# Effects of multiple edge cracks, shear force, elastic foundation, and boundary conditions on bucking of small-scale pillars[†]

Hossein Darban[a], Raimondo Luciano[b], Michał Basista[a1]

[a] *Institute of Fundamental Technological Research, Polish Academy of Sciences, Pawińskiego 5B, 02-106 Warsaw, Poland*

[b] *Department of Engineering, University of Naples Parthenope, 80133, Naples, Italy*

**Abstract**

The buckling instability of micro- and nanopillars can be an issue when designing intelligent miniaturized devices and characterizing composite materials reinforced with beam-like particles on the small-scale. Analytical modeling of the buckling of miniaturized pillars is especially important due to the difficulties in conducting experiments. Here, a well-posed stress-driven nonlocal model is developed, which allows the calculation of the critical loads and buckling configurations of the miniaturized pillars on an elastic foundation and with arbitrary numbers of edge cracks. The discontinuities in bending slopes and deflection at the damaged cross-sections due to the edge cracks are captured through the incorporation of both rotational and translational springs. A comprehensive analysis is conducted to investigate the instability of pillars containing a range of one to four cracks. This analysis reveals interesting effects regarding the influence of crack location, nonlocality, and elastic foundation on the initial and subsequent critical loads and associated buckling configurations. The main findings are: (i) the shielding and amplification effects related to a system of cracks become more significant as the dimensions of pillars reduce, (ii) the influence of the shear force at the damaged cross-section related to the translational spring must not be neglected when dealing with higher modes of buckling and long cracks, (iii) an elastic foundation decreases the effects of the cracks and size dependency on the buckling loads, and (iv) the effects of the edge cracks on the critical loads and buckling configurations of the miniaturized pillars are highly dependent on the boundary conditions.

**Keywords:** Miniaturized pillars; Edge crack; Spring model; Size effect; Instability.

# 1. Introduction

The buckling instability of the small-scale structural elements must be understood to safely design smart devices. Buckling experiments on the miniaturized specimens are difficult to conduct since applying appropriately aligned uniform compression loads at the specimen's ends is challenging. The atomistic models (e.g., [1-4]) are computationally expensive and therefore, cannot be used for design purposes. For these reasons, the engineering science community has always appreciated the analytical modeling of the buckling of micro- and nanosized pillars and beams. Such analytical knowledge is also important for material science applications since the buckling of beam-like reinforcements during the manufacturing process may significantly compromise the structural integrity and mechanical properties of composite materials such as carbon nanotube-reinforced polymer composites [5]. Using molecular dynamics simulations in [6] and a nonlocal analytical formulation in [7], it is shown that the buckling behavior of the micro- and nanopillars is highly affected by the presence of the edge cracks due to, for example, manufacturing process [8], in-service loadings, and environmental effects. Hence, the paper is aimed at developing an efficient beam formulation to model size-dependent buckling of miniaturized cracked pillars.

The formulation presented in [7] for buckling of the Bernoulli-Euler micro- and nanocantilevers with single edge cracks is extended here to miniaturized pillars supported by an elastic foundation and with any number of edge cracks and different kinematic constraints. Unlike the work in [7], the formulation takes into account the effect related to the shear force acting on the cracked cross-sections. All the improvements in the present formulation compared to that in [7] are important, as explained in the following. The consideration of the elastic foundation has practical application since the distributed lateral force applied to the pillar by the elastic foundation may be used to model the interaction between (i) the pillar and other elements in a miniaturized device, or (ii) a beam-like reinforcement and matrix phase in a composite material. Also, several edge cracks may appear in miniaturized structures in practical applications [9, 10], and therefore, the influence of a series of cracks on the buckling behavior of small-scale pillars must be studied. It will be proven in this work that excluding the impact of shear force would result in high errors for the miniaturized beams with severely damaged cross-sections and higher order buckling configurations. In addition, it will be shown here that the kinematic constraints at the pillar's end significantly influence the impact of the cracks on the critical loads of the miniaturized pillars. It is worth noting that the effects of crack propagation and interaction on the buckling behavior are not considered in this work. Also, only conservative forces are considered here and the related instability problems due to nonconservative forces are investigated elsewhere.

To solve the problem of buckling of a miniaturized pillar with multiple cracks, the structure is partitioned into distinct sections interconnected with rotational and translational springs. Therefore, the buckled configuration of the pillar is characterized by discontinuities in bending rotation and deflection where the cracks are located. The rotation and displacement jumps are assumed to be proportional to, respectively, the bending moment and the shear force at the cracked cross-section. The proportionality factors are the spring compliances defined solely in terms of the crack lengths through energy considerations and the linear elastic fracture mechanics principles.

It is widely acknowledged that the mechanical characteristics of materials and structures may become size-dependent at small scales [11, 12]. Several nonclassical continuum mechanics-based theories have been developed to address the size effect in the mechanical behavior of miniaturized structures. Among them, nonlocal elasticity and damage models (e.g., [13]) have gained popularity for analyzing the behavior of structures and materials. These models consider the long-range interactions between atoms and molecules, which are important at small scales and cannot be accounted for using local continuum mechanics models. Several studies have shown the applicability of nonlocal elasticity models in predicting the mechanical properties of various structures at the micro and nanoscale. As an example, in the widely referenced publication detailed in [14], several nonlocal beam formulations are developed utilizing Eringen's nonlocal theory [15] together with the Bernoulli-Euler, Timoshenko, and higher-order beam theories. These formulations are employed to predict the bending, vibration, and buckling of such beams. Similarly, nonlocal beam models are used in [16] to study wave propagation in carbon nanotubes, and in [17] to investigate the size-dependent buckling of beams. These studies, along with many others, demonstrate the potential of nonlocal elasticity models to provide reliable predictions of the mechanical behavior of miniaturized structures.

In this paper, the nonlocal stress-driven theory [18] is used to model the size effect. The stress-driven theory defines the elastic curvature at a section by taking into account the bending moment at all sections through an integral convolution, along with a smoothing kernel function that assigns weights to long-range interactions. A recent study in [19] demonstrated that the stress-driven model is capable of modeling size dependency in quasi-static and dynamic experiments conducted on micro- and nanocantilevers. The range of applicability of the stress-driven theory lies within materials that exhibit a stiffening behavior at small-scale dimensions, such as the polymer SU-8 as demonstrated by the experimental data reported in [20]. The stress-driven nonlocal model has proven effective in studying various problems concerning miniaturized structures, as demonstrated in prior works such as in [21-26]. These analytical investigations established that the nonlocal stress-driven theory can accurately simulate the mechanical behavior of

micro- and nanobeams, and the experimental confirmations outlined in [19] enhance the reliability and usefulness of the model.

Following previous studies[7, 27, 28] and with the assumption of an exponential kernel function, integral formulation of the nonlocal constitutive expression, which is defined over the entire length of the pillar, is converted into a differential equation valid at different sections of the pillar between damaged cross-sections. The associated higher-order constitutive boundary and continuity conditions are also derived. These higher-order conditions are expressed in terms of curvature. Therefore, the formulated model is always well-posed and does not encounter inconsistencies such as those that occur when the models based on Eringen's nonlocal theory [15] are used (see, for example, [29]). The problem definition, the main assumptions, and the model derivation are detailed in Sect. 2. Section 3 applies the formulation to pillars with one, two, three, and four damaged cross-sections, and extensively studies the effects of damage location and severity, size-dependency, and elastic foundation on the fundamental and higher-order buckling loads and configurations. In Sect. 4 a wrap-up and conclusions of this research are presented.

## 2. Definition and Formulation of the Problem

The system to be analyzed is depicted in Fig. 1. It involves a slender pillar of micro- and nanoscale dimensions being subjected to a compressive load, denoted as $P$. The pillar has $n$ edge cracks and is resting on an elastic foundation. The length of the pillar is $L$, and its cross-section is rectangular, with in-plane thickness $h$ and out-of-plane width $b$. The pillar is made of a homogeneous and isotropic material and deforms under plane stress conditions. The size-dependent critical loads and buckling mode shapes of the pillar are sought. The mid-thickness and lower end of the pillar serve as the origin of the Cartesian coordinate system $x - y$, used for the formulation of the problem. The cracks are numbered from bottom to top, located at $x_i$, and have different lengths $a_i$ for $i = 1, \ldots, n$.

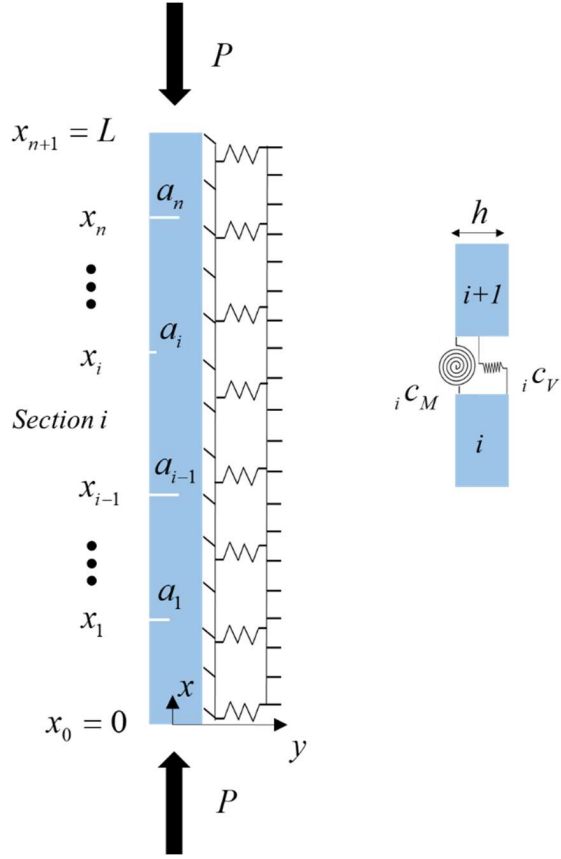

**Fig. 1** A miniaturized pillar with *n* edge cracks supported by an elastic foundation under the action of the compressive load *P*. Rotational and translational springs are used to model the influence of cracks [30-37].

## 2.1. Formulation

### 2.1.1 Crack Kinematic Compatibility Conditions

To include the cracks in the proposed model, it is necessary to decompose the pillar into $n+1$ sections at the damaged cross-sections. The sections, numbered from bottom to top, are assumed to be connected by rotational and translational springs, as shown in Fig. 1 [30-37]. As the pillar undergoes buckling, it is reasonable to assume that the configuration resulting from compressive forces causes the edge cracks within the pillar to open up. Therefore, the bending rotation and the deflection of the deformed pillar have jumps at the cracked cross-sections, $x_i$ for $i = 1, \ldots, n$:

$$^{(i+1)}v_{,x}(x_i) - {}^{(i)}v_{,x}(x_i) = {}_ic_M M(x_i)$$
$$^{(i+1)}v(x_i) - {}^{(i)}v(x_i) = {}_ic_V V(x_i) \tag{1}$$

Here, $^{(i)}v$ denotes the deflection (i.e. transverse displacement) of section $i$, and $^{(i)}v_{,x}$ represents its first derivative (i.e. the bending slope). When a quantity displays a superscript $(i)$ on the left, it signifies its association with section $i$, and $F_{,\underbrace{xxx...x}_{j}} = d^j F/dx^j$. $M(x_i)$ and $V(x_i)$ represent the bending moment and shear force at the damaged cross-section, respectively. Additionally, $_iC_M$ and $_iC_V$ stand for the compliances of the rotational and translational springs. These parameters directly relate to the stress intensity factors of the opening and sliding modes arising from the bending moment and shear force. Under the assumption that the stress field of each crack is unaffected by the presence of the other cracks, the compliances of the rotational and translational springs are derived in [33] and presented in Appendix A.

*2.1.2 Nonlocal Formulation*

The stress-driven nonlocal theory of nonlocal elasticity [18] is used to capture the small-scale size effect in the buckling behavior of the miniaturized pillar. Within the context of the stress-driven theory, the pointwise strain is defined through a convolution integral that includes the stresses exerted at all points within the body and the corresponding kernel function. Integrating the constitutive equation over the thickness, the following constitutive equation is derived for the pillar [18]:

$$\chi(x) = \int_0^L \frac{1}{2L_C} e^{\left(-\frac{|x-\xi|}{L_C}\right)} CM(\xi) d\xi \qquad (2)$$

where the elastic compliance is $C = 12/(Ebh^3)$, with $E$ being Young's modulus. The relation in Eq. (2) defines the curvature $\chi$ as an integral convolution over the length of the pillar in terms of the bending moment $M$ and the kernel $\phi_{L_C}(x) = 0.5 e^{(-|x|/L_C)}/L_C$. The length scale parameter $L_C$ determines the characteristics of the kernel function. For larger values of the length scale parameter, the kernel function exhibits a smoother behavior. When $L_C \to 0$, the kernel function in Eq. (2) becomes nonzero only at $x$ and the constitutive equation reduces to that of the local formulation.

The constitutive equation (2) is defined over the entire length of the pillar, and therefore, can be used to solve only the buckling problem of the intact pillar [21, 22]. Following [7, 27, 28] and for the pillar with $n$ edge cracks, the constitutive equation (2) can be expressed as:

$$\chi(x) = \sum_{i=1}^{n+1} \int_{x_{i-1}}^{x_i} \frac{1}{2L_C} e^{\left(-\frac{|x-\xi|}{L_C}\right)} C^{(i)} M(\xi) d\xi \tag{3}$$

The curvature of the generic section $i$ at the distance $x$ has four different contributions based on Eq. (3) from the cross-sections within $0 \leq \xi \leq x_{i-1}$, $x_{i-1} \leq \xi \leq x$, $x \leq \xi \leq x_i$, and $x_i \leq \xi \leq L$. Using this decomposition technique, taking the first and second derivatives of the constitutive equation, and after some manipulations, the integral equation (2) can be written in the following differential form:

$$^{(i)}\chi - L_C^2 {}^{(i)}\chi_{,xx} = C^{(i)} M \tag{4}$$

for $i = 1, \ldots, n+1$, subject to the constitutive boundary and continuity conditions:

$$^{(i)}\chi_{,x}(x_{i-1}) = \frac{1}{L_C} \left[ {}^{(i)}\chi(x_{i-1}) - \sum_{k=1}^{i-1} \int_{x_{k-1}}^{x_k} \left( \frac{1}{L_C} e^{\frac{\xi - x_{i-1}}{L_C}} C^{(k)} M(\xi) \right) d\xi \right]$$

$$^{(i)}\chi_{,x}(x_i) = -\frac{1}{L_C} \left[ {}^{(i)}\chi(x_i) - \sum_{k=i+1}^{n+1} \int_{x_{k-1}}^{x_k} \left( \frac{1}{L_C} e^{\frac{x_i - \xi}{L_C}} C^{(k)} M(\xi) \right) d\xi \right] \tag{5}$$

for $i = 1, \ldots, n+1$. Note that the summation is equal to zero when the upper bound is less than the lower bound. The nonlocal constitutive equation of the pillar presented in Eq. (4) is two orders higher than that of the local formulation. Therefore, for a miniaturized pillar with $n$ edge cracks, the formulation will be $2(n+1)$ orders higher than the local formulation. The well-conditioning of the nonlocal formulation is guaranteed by the $2(n+1)$ higher-order constitutive boundary and continuity conditions in Eq. (5).

The curvature in Eqs. (4) and (5) is related to the transverse displacement through the relationship between curvature and deflection in the Bernoulli-Euler beam model, $^{(i)}\chi = {}^{(i)}v_{,xx}$. Therefore, the constitutive equations (4) and (5) expressed through the deflection are:

$$^{(i)}v_{,xx} - L_C^2 {}^{(i)}v_{,xxxx} = C^{(i)} M \tag{6}$$

$$^{(i)}v_{,xxx}(x_{i-1}) = \frac{1}{L_C}\left[ {}^{(i)}v_{,xx}(x_{i-1}) - \sum_{k=1}^{i-1}\int_{x_{k-1}}^{x_k}\left(\frac{1}{L_C}e^{\frac{\xi-x_{i-1}}{L_C}}\left[{}^{(k)}v_{,xx}(\xi) - L_C^2\,{}^{(k)}v_{,xxxx}(\xi)\right]\right)d\xi\right]$$

$$^{(i)}v_{,xxx}(x_i) = -\frac{1}{L_C}\left[ {}^{(i)}v_{,xx}(x_i) - \sum_{k=i+1}^{n+1}\int_{x_{k-1}}^{x_k}\left(\frac{1}{L_C}e^{\frac{x_i-\xi}{L_C}}\left[{}^{(k)}v_{,xx}(\xi) - L_C^2\,{}^{(k)}v_{,xxxx}(\xi)\right]\right)d\xi\right] \quad (7)$$

for $i = 1, \ldots, n+1$.

### 2.1.3 Variationally Consistent Equations

On the basis of the Bernoulli-Euler beam model, the buckling of each section of the pillar is governed by the differential equation $^{(i)}M_{,xx}(x) + P^{(i)}v_{,xx}(x) - {}^{(i)}q(x) = 0$. The last term in this equation, $^{(i)}q(x)$, represents the force distribution resulting from the elastic foundation acting on the pillar's section. A two-parameter Pasternak elastic foundation, which defines the response of the foundation by $^{(i)}q(x) = -\left(K_W\,{}^{(i)}v(x) - K_P\,{}^{(i)}v_{,xx}(x)\right)$, is assumed here. $K_W$ and $K_P$ are the Winkler and Pasternak parameters. The buckling equation at each section of the pillar can be expressed as function of the deflection through the utilization of the constitutive equation (6), namely:

$$L_C^2\,{}^{(i)}v_{,xxxxxx} - {}^{(i)}v_{,xxxx} - (P - K_P)C\,{}^{(i)}v_{,xx}(x) - CK_W\,{}^{(i)}v(x) = 0 \quad (8)$$

for $i = 1, \ldots, n+1$. The sixth-order differential equations (8) governing the buckling of the sections of the pillar deviate from the fourth-order equation found in the local Bernoulli-Euler theory. As a result, the importance of incorporating extra constitutive boundary and continuity conditions becomes apparent. The variationally consistent boundary conditions for typical cases are presented in Appendix B. The variationally consistent continuity conditions at the damaged cross-sections are:

$$^{(i+1)}v_{,x}(x_i) - {}^{(i)}v_{,x}(x_i) = {}_i c_M \frac{{}^{(i+1)}v_{,xx}(x_i) - L_C^2 \, {}^{(i+1)}v_{,xxxx}(x_i)}{C}$$

$$^{(i+1)}v(x_i) - {}^{(i)}v(x_i) = -{}_i c_V \left[ \frac{{}^{(i+1)}v_{,xxx}(x_i) - L_C^2 \, {}^{(i+1)}v_{,xxxxx}(x_i)}{C} + (P - K_P) \, {}^{(i+1)}v_{,x}(x_i) \right]$$

$$^{(i)}v_{,xx}(x_i) - L_C^2 \, {}^{(i)}v_{,xxxx}(x_i) = {}^{(i+1)}v_{,xx}(x_i) - L_C^2 \, {}^{(i+1)}v_{,xxxx}(x_i) \qquad (9)$$

$$\frac{{}^{(i)}v_{,xxx}(x_i) - L_C^2 \, {}^{(i)}v_{,xxxxx}(x_i)}{C} + (P - K_P) \, {}^{(i)}v_{,x}(x_i) =$$

$$\frac{{}^{(i+1)}v_{,xxx}(x_i) - L_C^2 \, {}^{(i+1)}v_{,xxxxx}(x_i)}{C} + (P - K_P) \, {}^{(i+1)}v_{,x}(x_i)$$

for $i = 1, \ldots, n$. The first two equations address the kinematic compatibility requirements at the damaged cross-section, as outlined in Eq. (1). The third and fourth equations, on the other hand, ensure continuous transitions of the bending moment and shear force at the damaged cross-sections.

## 2.2. Non-Dimensional Forms of Equations

To simplify the mathematical formulation of the problem, the subsequent equations are expressed using the following dimensionless parameters:

$$^{(i)}\eta = \frac{{}^{(i)}v}{L}; \quad \bar{x} = \frac{x}{L}; \quad \bar{h} = \frac{h}{L}; \quad \alpha = \frac{PL^2 C}{\pi^2}; \quad {}_i \bar{c}_M = \frac{{}_i c_M}{CL}; \quad {}_i \bar{c}_V = \frac{{}_i c_V}{CL^3}; \quad \lambda = \frac{L_C}{L} \qquad (10)$$

$$\bar{K}_P = K_P L^2 C; \quad \bar{K}_W = K_W L^4 C$$

Based on the dimensionless parameters, the buckling equation (8) and the continuity conditions in Eq. (9) are:

$$\lambda^2 \, {}^{(i)}\eta_{,\bar{x}\bar{x}\bar{x}\bar{x}\bar{x}\bar{x}}(\bar{x}) - {}^{(i)}\eta_{,\bar{x}\bar{x}\bar{x}\bar{x}}(\bar{x}) - \left(\alpha \pi^2 - \bar{K}_P\right) {}^{(i)}\eta_{,\bar{x}\bar{x}}(\bar{x}) - \bar{K}_W \, {}^{(i)}\eta(\bar{x}) = 0 \qquad (11)$$

$$^{(i+1)}\eta_{,\bar{x}}(\bar{x}_i) - {}^{(i)}\eta_{,\bar{x}}(\bar{x}_i) = {}_i \bar{c}_M \left[ {}^{(i+1)}\eta_{,\bar{x}\bar{x}}(\bar{x}_i) - \lambda^2 \, {}^{(i+1)}\eta_{,\bar{x}\bar{x}\bar{x}\bar{x}}(\bar{x}_i) \right]$$

$$^{(i+1)}\eta(\bar{x}_i) - {}^{(i)}\eta(\bar{x}_i) = -{}_i \bar{c}_V \left[ {}^{(i+1)}\eta_{,\bar{x}\bar{x}\bar{x}}(\bar{x}_i) - \lambda^2 \, {}^{(i+1)}\eta_{,\bar{x}\bar{x}\bar{x}\bar{x}\bar{x}}(\bar{x}_i) + \left(\alpha \pi^2 - \bar{K}_P\right) {}^{(i+1)}\eta_{,\bar{x}}(\bar{x}_i) \right]$$

$$^{(i)}\eta_{,\bar{x}\bar{x}}(\bar{x}_i) - \lambda^2 \, {}^{(i)}\eta_{,\bar{x}\bar{x}\bar{x}\bar{x}}(\bar{x}_i) = {}^{(i+1)}\eta_{,\bar{x}\bar{x}}(\bar{x}_i) - \lambda^2 \, {}^{(i+1)}\eta_{,\bar{x}\bar{x}\bar{x}\bar{x}}(\bar{x}_i) \qquad (12)$$

$$\left[ {}^{(i)}\eta_{,\bar{x}\bar{x}\bar{x}}(\bar{x}_i) - \lambda^2 \, {}^{(i)}\eta_{,\bar{x}\bar{x}\bar{x}\bar{x}\bar{x}}(\bar{x}_i) \right] + \left(\alpha \pi^2 - \bar{K}_P\right) {}^{(i)}\eta_{,\bar{x}}(\bar{x}_i) =$$

$$\left[ {}^{(i+1)}\eta_{,\bar{x}\bar{x}\bar{x}}(\bar{x}_i) - \lambda^2 \, {}^{(i+1)}\eta_{,\bar{x}\bar{x}\bar{x}\bar{x}\bar{x}}(\bar{x}_i) \right] + \left(\alpha \pi^2 - \bar{K}_P\right) {}^{(i+1)}\eta_{,\bar{x}}(\bar{x}_i)$$

Table B1 (Appendix B) displays the dimensionless variational boundary conditions, while the dimensionless form of Eq. (7) is:

$$^{(i)}\eta_{,\overline{xxx}}(\overline{x}_{i-1}) = \frac{1}{\lambda}\left[^{(i)}\eta_{,\overline{xx}}(\overline{x}_{i-1}) - \sum_{k=1}^{i-1}\int_{\overline{x}_{k-1}}^{\overline{x}_k}\left(\frac{1}{\lambda}e^{\frac{\xi-\overline{x}_{i-1}}{\lambda}}\left[^{(k)}\eta_{,\xi\xi}(\xi) - \lambda^2\,^{(k)}\eta_{,\xi\xi\xi\xi}(\xi)\right]\right)d\xi\right]$$

$$^{(i)}\eta_{,\overline{xxx}}(\overline{x}_i) = -\frac{1}{\lambda}\left[^{(i)}\eta_{,\overline{xx}}(\overline{x}_i) - \sum_{k=i+1}^{n+1}\int_{\overline{x}_{k-1}}^{\overline{x}_k}\left(\frac{1}{\lambda}e^{\frac{\overline{x}_i-\xi}{\lambda}}\left[^{(k)}\eta_{,\xi\xi}(\xi) - \lambda^2\,^{(k)}\eta_{,\xi\xi\xi\xi}(\xi)\right]\right)d\xi\right]$$

(13)

The size-dependent buckling equation (11) is a linear homogeneous sixth-order ordinary differential equation with constant coefficients whose closed-form solution is derived in [22] in terms of six unknown constants. For the pillar with $n$ cracks, the buckled configuration is defined in terms of $6\times(n+1)$ unknown constants that must satisfy the $4\times n$ variationally consistent continuity conditions in Eq. (12), 4 boundary conditions given in Appendix B, and $2\times(n+1)$ constitutive boundary and continuity conditions in Eq. (13). The critical loads and associated buckling configurations are determined by solving the resulting eigenvalue problem.

## 3. Numerical Examples and Discussion

This section presents the buckling loads and mode shapes of pillars with and without elastic foundation with one to four cracks. The first three modes of buckling and different boundary conditions are considered and the critical loads are given by varying the crack length, crack location, the parameters of the elastic foundation, and nonlocality. The bisection method is used to numerically calculate the roots of the determinant of the coefficient matrix, which correspond to the buckling loads.

### 3.1. Verification

To assess the accuracy of the proposed formulation, the critical loads predicted by the model are compared with previously reported results for three distinct cases. The first case is depicted in Fig. 2, where the critical loads associated with a small-scale pillar containing one crack of length of $a/h$ located at the middle of the pillar are presented. The normalized buckling loads are determined with respect to the buckling loads of the undamaged pillars calculated in [22].

The results refer to the case with $\lambda = 0.25$, $h/L = 0.05$, $\bar{K}_W = 400$ and $\bar{K}_P = 8$. Four different boundary conditions are considered. The outcomes are demonstrated through decreasing the length of the crack. It is evident form Fig. 2 that the predictions of the proposed model are in agreement with those reported in [22] as the crack length approaches zero, i.e., in the limit of $h/a \to \infty$.

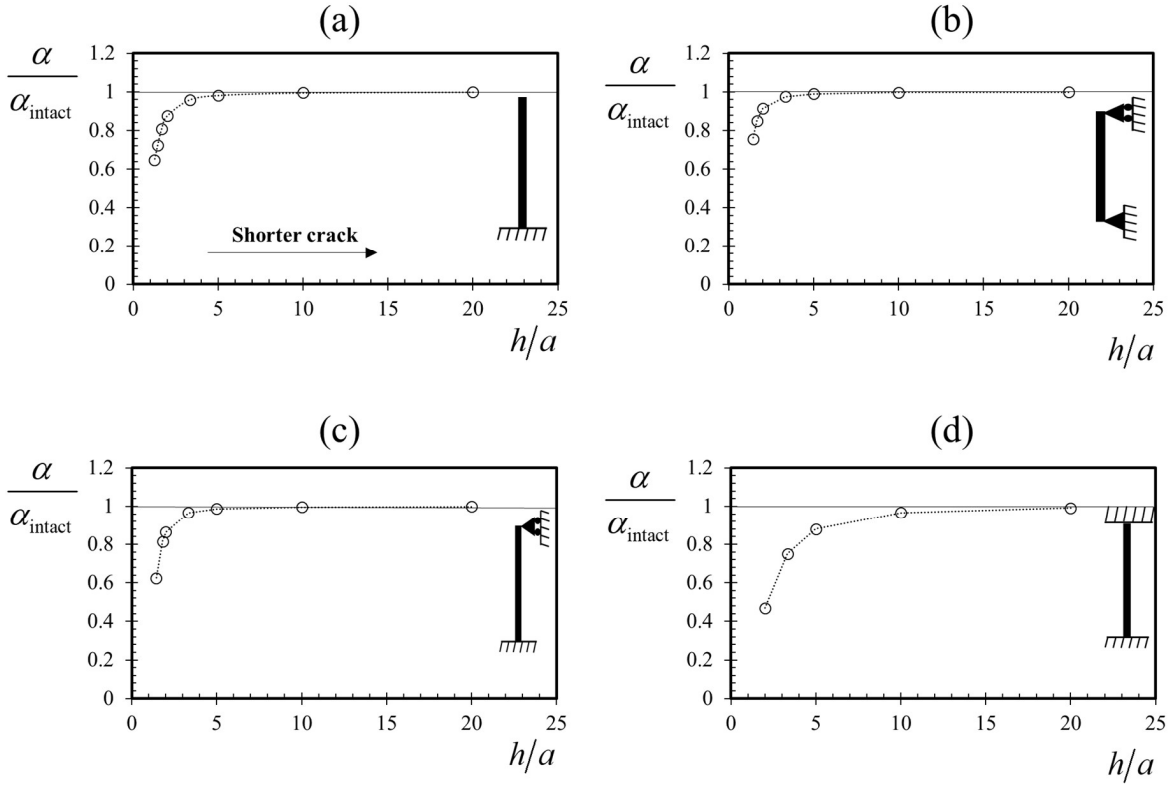

**Fig. 2** Ratios of the buckling loads upon reducing the crack length of: (a) clamped-free, (b) pinned-pinned, (c) clamped-pinned, and (d) clamped-clamped miniaturized pillars with one crack at the pillar's mid-length. The results are normalized against the solutions of the intact pillars and presented for $\lambda = 0.25$, $h/L = 0.05$, $\bar{K}_W = 400$ and $\bar{K}_P = 8$.

The results depicted in Fig. 3 refer to the buckling loads of pillars with one damage at the mid-length with the severity of $a/h = 0.31$, and $h/L = 0.05$. The elastic foundation is absent in this example. The results are normalized against the results obtained in [38] using the local Bernoulli-Euler beam model and presented for different boundary conditions by reducing the nonlocal parameter. For all the cases, the results of the model ultimately align with the data reported in [38] as the nonlocal parameter decreases.

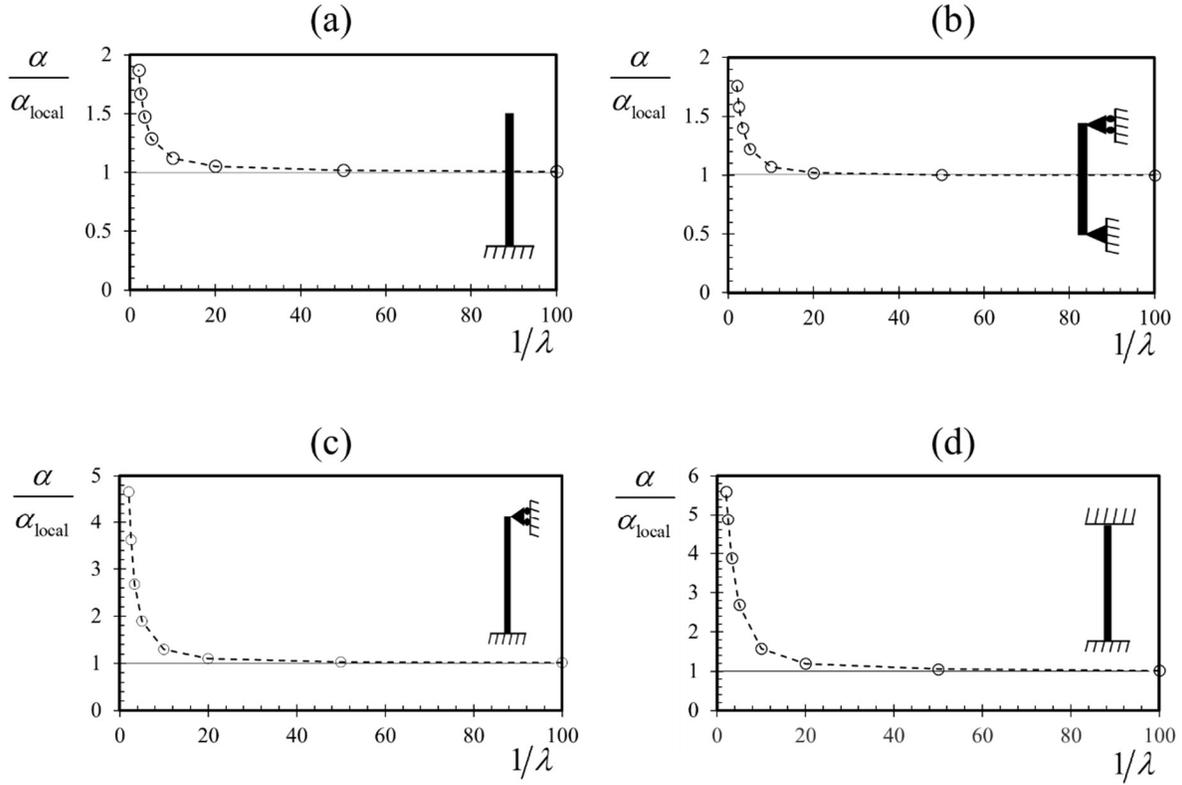

**Fig. 3** Ratios of the buckling loads vs. the nonlocal parameter for (a) clamped-free, (b) pinned-pinned, (c) clamped-pinned, and (d) clamped-clamped miniaturized pillars without elastic foundation with one crack at the mid-length. The results are normalized against the solutions of the local pillars for $a/h = 0.31$, $h/L = 0.05$.

The third case which is used to verify the present formulation is a miniaturized cantilever with two cracks at $\bar{x}_1 = 0.3$ and $\bar{x}_2 = 0.5$, $h/L = 0.05$, $\lambda = 0.5$, and $\bar{K}_W = \bar{K}_P = 0$. In the calculations, the length of the second crack is kept constant and equal to $a_2/h = 0.4$. The first three critical loads are presented in Fig. 4 by reducing the length of the first crack. In the limit when the length of the first crack goes to zero, i.e. $h/a_1 \to \infty$, the buckling loads determined by the present formulation tend to those obtained in [7] for the cantilevers with one damaged cross-section located at the middle of the structure.

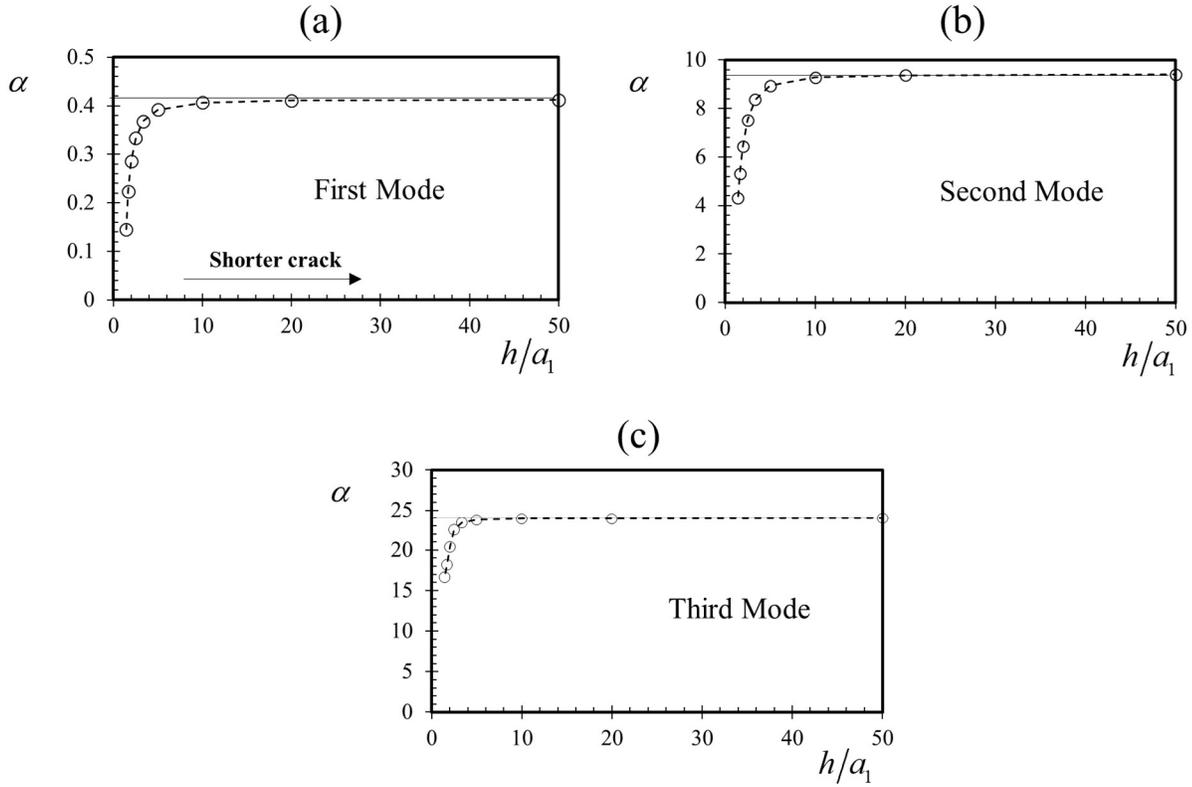

**Fig. 4** The dimensionless buckling loads of a cantilever pillar in the absence of the elastic foundation with two cracks at $\bar{x}_1 = 0.3$ and $\bar{x}_2 = 0.5$, $a_2/h = 0.4$, $h/L = 0.05$, and $\lambda = 0.5$ by varying $a_1/h$. Results are presented for the (a) first, (b) second, and (c) third modes. The results tend to those given in [7] for the cantilever with one crack at the mid-length as the length of the first crack goes to zero, $h/a_1 \to \infty$.

### 3.2. Miniaturized Pillars with One Crack

*3.2.1 Effects of Nonlocality, Boundary Conditions, and Elastic Foundation*

In [7], the critical loads of micro- and nanocantilevers containing a single crack are presented. Here, the results are extended to the pillars with pinned-pinned, clamped-pinned, and clamped-clamped boundary conditions to investigate how the kinematic constraints would change the effect of the crack on the buckling loads. For this purpose, a pillar with one crack at the mid-length, $h/L = 0.05$, and $\bar{K}_W = \bar{K}_P = 0$ is considered. The buckling loads are normalized against the solutions of the intact pillars and illustrated in Fig. 5 by changing the crack length for different nonlocal parameters.

For all the cases, the existence of a longer crack reduces the stiffness of the pillars and therefore, the buckling loads decrease. Since the clamped-clamped pillar is stiffer than the pillars with other types of boundary conditions, the presence of the mid-length crack reduces its stiffness to a greater extent. The is the underlying reason for the high dependency of the buckling loads in clamped-clamped pillars on the crack presence. In addition, the impact of the crack on the critical loads is stronger for the nonlocal pillars. For illustration, the crack with $a/h = 0.3$ causes a 9, 12, and 17% reduction in the buckling load of the pinned-pinned pillar with $\lambda = 0$ 0.3, and 0.6. Or, a relatively short crack of $a/h = 0.1$ at the mid-length of the miniaturized clamped-clamped pillar ( $\lambda = 0.5$ ) causes a 12% reduction in the critical load, while this decrease for a large-scale pillar ( $\lambda = 0$ ) is only 1%.

As a result, the influence of damage on buckling loads is contingent on both the kinematic constraints at the pillar's ends and the value assigned to the nonlocal parameter. Based on this observation, one can conclude that the occurrence of defects in the form of edge cracks is a more serious issue in the design of the miniaturized pillars than that of the large-scale pillars.

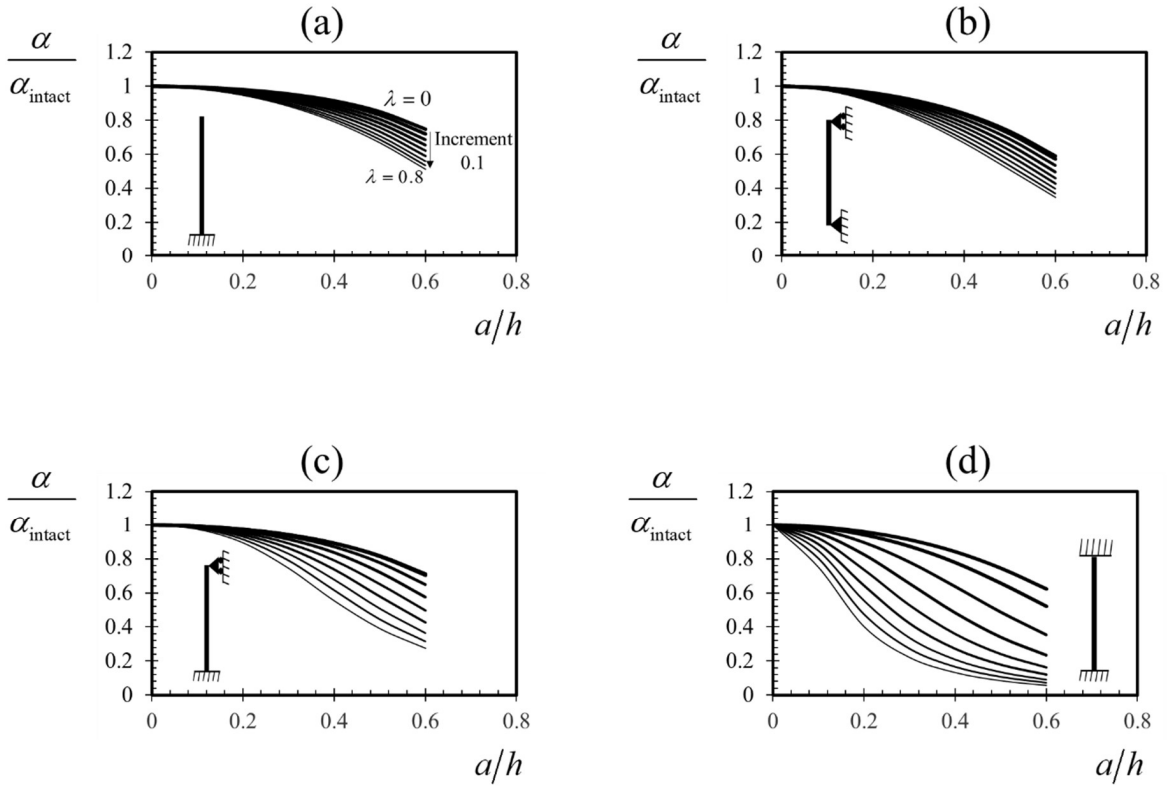

**Fig. 5** The critical loads of (a) clamped-free, (b) pinned-pinned, (c) clamped-pinned, and (d) clamped-clamped pillars in the absence of the elastic foundation by varying the nonlocal parameter and crack length. Pillars have one crack at the mid-length, and $h/L = 0.05$.

To analyze how the buckling loads of the cracked pillars are affected by the elastic foundation, the previous example is considered in the presence of an elastic foundation with $\bar{K}_W = 200$ and $\bar{K}_P = 6$, and the results are presented in Fig. 6. It can be understood from the comparison of Fig. 6 with Fig. 5 that the buckling loads of the pillars in the presence of the elastic foundation are generally less affected by the crack length. This observation can be explained by recalling the fact that the global stiffness of the system is composed of two contributions: (i) the stiffness of the pillar, and (ii) the stiffness of the elastic foundation. While the former is reduced due to the presence of the crack, the latter is unaffected by the crack. Therefore, the crack has a weaker effect on the buckling loads of pillars supported by the elastic foundation. As an illustration, critical load ratios of pinned-pinned pillars with and without cracks for $a/h = 0.5$ and $\lambda = 0.5$ are 0.82 and 0.59 in the presence and absence of the elastic foundation, respectively.

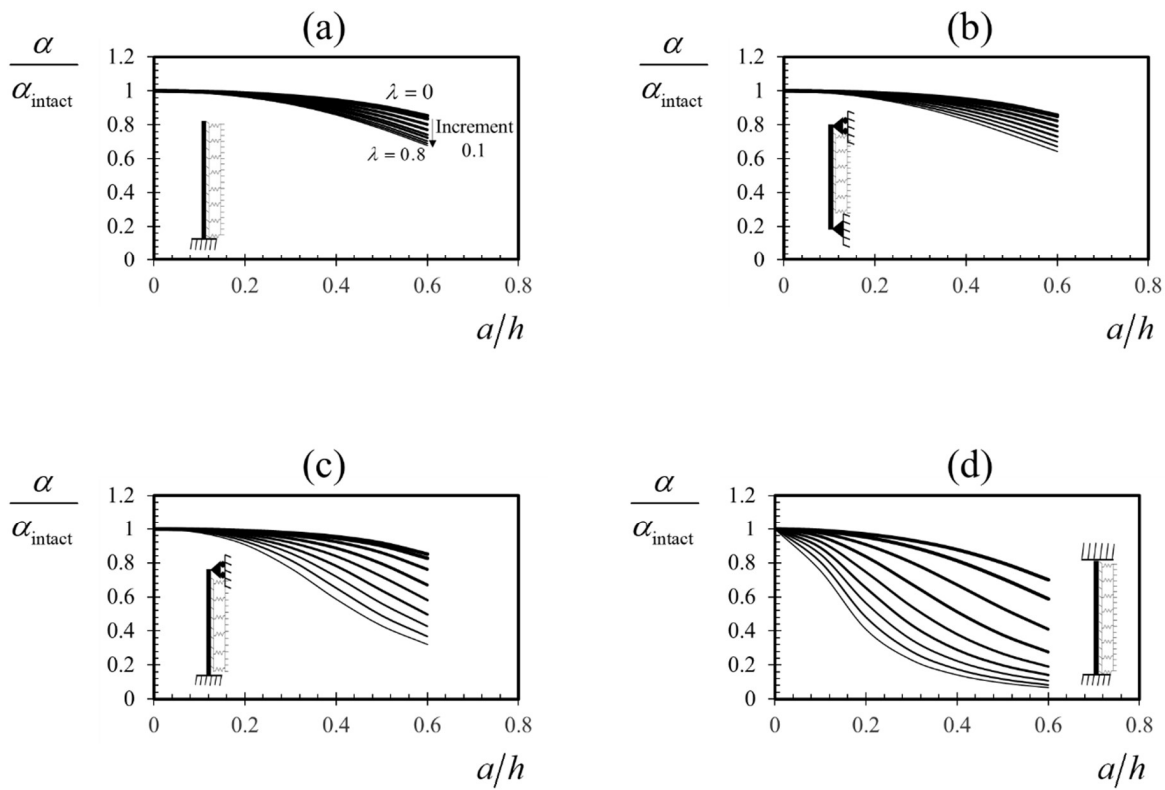

**Fig. 6** The critical loads as a function of the nonlocality and crack length for (a) clamped-free, (b) pinned-pinned, (c) clamped-pinned, and (d) clamped-clamped pillars supported by an elastic foundation with $\bar{K}_W = 200$ and $\bar{K}_P = 6$. Pillars have one crack at the mid-length, and $h/L = 0.05$.

The critical loads of pillars are influenced by the elastic foundation differently according to the boundary conditions imposed. Generally, the elastic foundation has a stronger effect on the buckling loads of less constrained pillars. To verify this statement, the buckling loads of intact and cracked pillars with and without elastic foundation are given in Table 1 for typical boundary conditions. Pillars have one crack at the mid-span, and $h/L = 0.05$. The elastic foundation increases the critical loads of pillars with $a/h = 0.2$ and $\lambda = 0.2$, by 731, 223, 57, and 18%, for the clamped-free, pinned-pinned, clamped-pinned, and clamped-clamped boundary conditions. These percentage changes for the pillar with $a/h = 0$ (intact pillar) are 725, 213, 55, and 16%, and for the pillar with $a/h$ equal to 0.4, are 758, 262, 64, and 22%. Therefore, the influence of the elastic foundation on the critical loads does not significantly change for the intact and cracked pillars considered in Table 1.

**Table 1:** Critical loads of miniaturized intact and cracked pillars in the absence and presence of the elastic foundation. Pillars have one crack at the mid-length, and $h/L = 0.05$.

| $\bar{K}_W = \bar{K}_P = 0$ | | | | | | | | | |
|---|---|---|---|---|---|---|---|---|---|
| Boundary Conditions | $a/h = 0$ | | | $a/h = 0.2$ | | | $a/h = 0.4$ | | |
| | $\lambda = 0$ | $\lambda = 0.2$ | $\lambda = 0.4$ | $\lambda = 0$ | $\lambda = 0.2$ | $\lambda = 0.4$ | $\lambda = 0$ | $\lambda = 0.2$ | $\lambda = 0.4$ |
| Clamped-Free | 0.2500 | 0.3278 | 0.4348 | 0.2450 | 0.3189 | 0.4192 | 0.2286 | 0.2908 | 0.3715 |
| Pinned-Pinned | 1.0000 | 1.2505 | 1.6786 | 0.9605 | 1.1902 | 1.5732 | 0.8400 | 1.0117 | 1.2752 |
| Clamped-Pinned | 2.0457 | 3.9230 | 7.8727 | 1.9934 | 3.8055 | 7.5088 | 1.8259 | 3.4148 | 6.2346 |
| Clamped-Clamped | 4.0000 | 12.7439 | 34.7539 | 3.8421 | 11.4143 | 25.6152 | 3.3693 | 8.1243 | 12.5023 |
| $\bar{K}_W = 200, \bar{K}_P = 6$ | | | | | | | | | |
| Boundary Conditions | $a/h = 0$ | | | $a/h = 0.2$ | | | $a/h = 0.4$ | | |
| | $\lambda = 0$ | $\lambda = 0.2$ | $\lambda = 0.4$ | $\lambda = 0$ | $\lambda = 0.2$ | $\lambda = 0.4$ | $\lambda = 0$ | $\lambda = 0.2$ | $\lambda = 0.4$ |
| Clamped-Free | 2.1928 | 2.7013 | 3.3229 | 2.1645 | 2.6514 | 3.2380 | 2.0752 | 2.4963 | 2.9778 |
| Pinned-Pinned | 3.6611 | 3.9090 | 4.3315 | 3.6207 | 3.8495 | 4.2290 | 3.4848 | 3.6581 | 3.9147 |
| Clamped-Pinned | 4.2663 | 6.0637 | 9.9946 | 4.2214 | 5.9628 | 9.6622 | 4.0692 | 5.5959 | 8.3913 |
| Clamped-Clamped | 6.1180 | 14.8237 | 36.8620 | 5.9352 | 13.4344 | 27.5455 | 5.3700 | 9.9277 | 14.0802 |

The curves in Fig. 7 show the variations of the buckling loads of miniaturized intact and cracked pillars with one crack at $\bar{x}_1 = 0.3$ and $h/L = 0.05$. The data refer to the pillars with different boundary conditions without the elastic foundation. As shown in Fig. 7, the critical loads are always higher for pillars with higher nonlocality, regardless of the boundary conditions and the crack length. However, the nonlocal parameter has a less significant effect on the buckling loads of the cracked pillars than the intact pillars. For instance, the critical loads of the intact and cracked ($a/h = 0.6$) clamped-free pillars with $\lambda = 0.4$ are 74 and 39% higher than those of the local ($\lambda = 0$) pillars.

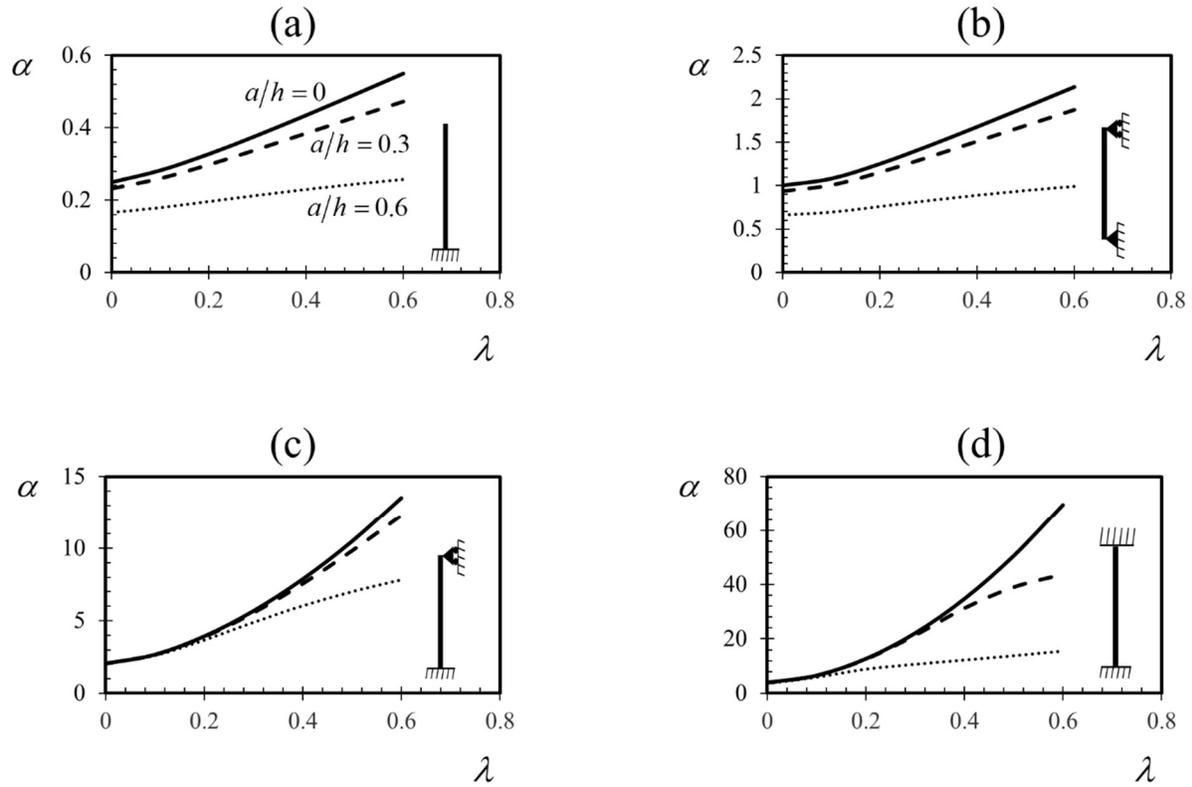

**Fig. 7** The critical loads of (a) clamped-free, (b) pinned-pinned, (c) clamped-pinned, and (d) clamped-clamped pillars without elastic foundation by varying the nonlocality parameter. Pillars have one damaged cross-section with $\bar{x}_1 = 0.3$, and $h/L = 0.05$.

To investigate the impact of the nonlocality on the critical loads of pillars in the presence of the elastic foundation, the same results as those illustrated in Fig. 7 are presented in Fig. 8 for $\bar{K}_W = 300$ and $\bar{K}_P = 8$. The influence of nonlocality on the critical loads across all four boundary conditions and three distinct crack lengths is moderated by the presence of an elastic foundation. Specifically, shifting the nonlocal parameter from 0 (local model) to 0.6 results in a 100% and 16% increase in the buckling loads in cases without and with an elastic foundation, respectively.

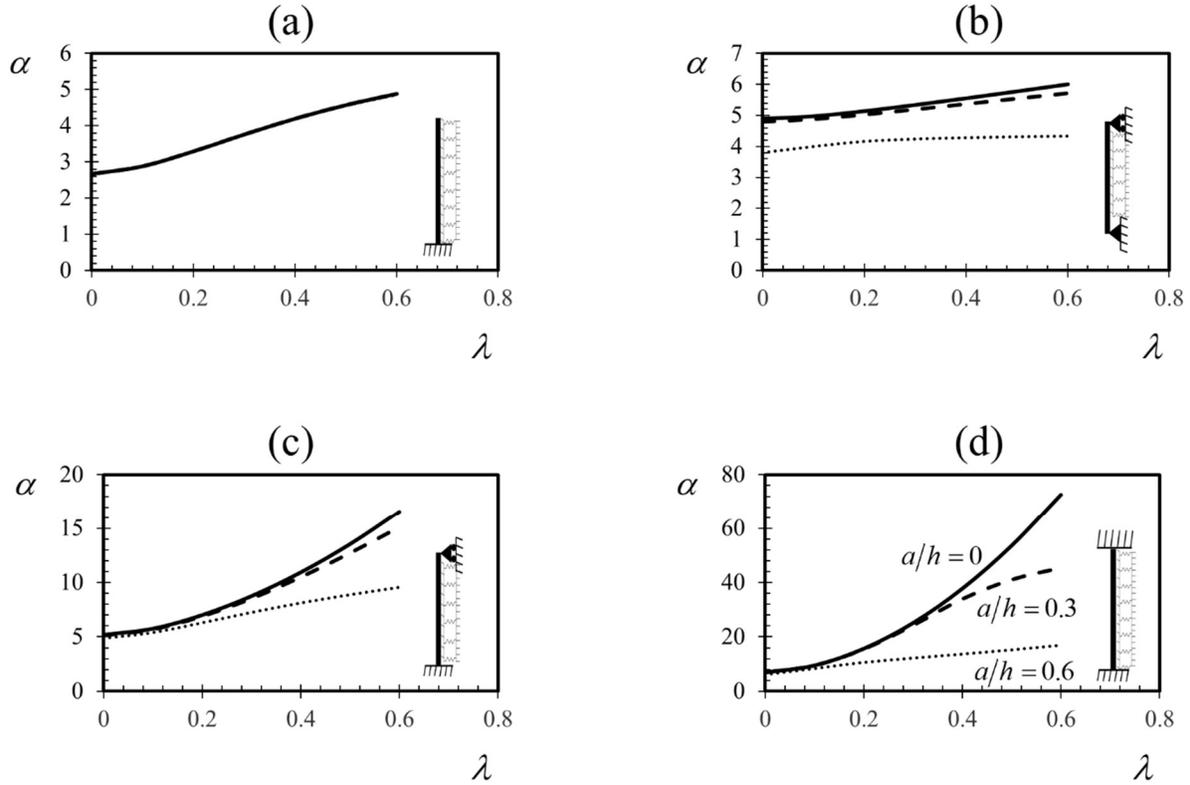

**Fig. 8** The critical loads vs. the nonlocality parameter of (a) clamped-free, (b) pinned-pinned, (c) clamped-pinned, and (d) clamped-clamped pillars supported by an elastic foundation with $\bar{K}_W = 300$ and $\bar{K}_P = 8$. Pillars have one damaged cross-section with $\bar{x}_1 = 0.3$, and $h/L = 0.05$.

An important understanding can be gained from Fig. 8(a), where the curves corresponding to the different crack lengths virtually coincide. This means that the crack has almost no effect on the buckling load of the clamped-free pillar supported by the elastic foundation. To investigate this observation, the first three buckling loads of the clamped-free pillar with $\lambda = 0.5$ are demonstrated in Fig. 9(a)-(c) by varying the crack length. The curve in Fig. 9(a) shows that the first buckling load is almost independent of the crack length. The axial variation of the normalized bending moment is illustrated in Fig. 9(d) for the clamped-free pillar without cracks, with consideration of the first, second, and third modes of buckling. The bending moment of the first mode of buckling is almost zero at $\bar{x} = 0.3$. Hence, the first buckling load becomes almost independent of the length of the crack located at $\bar{x} = 0.3$. However, the bending moments of the higher modes are not negligible at $\bar{x} = 0.3$, and therefore, the critical loads are highly dependent on the crack severity (see Fig. 9(b) and (c)).

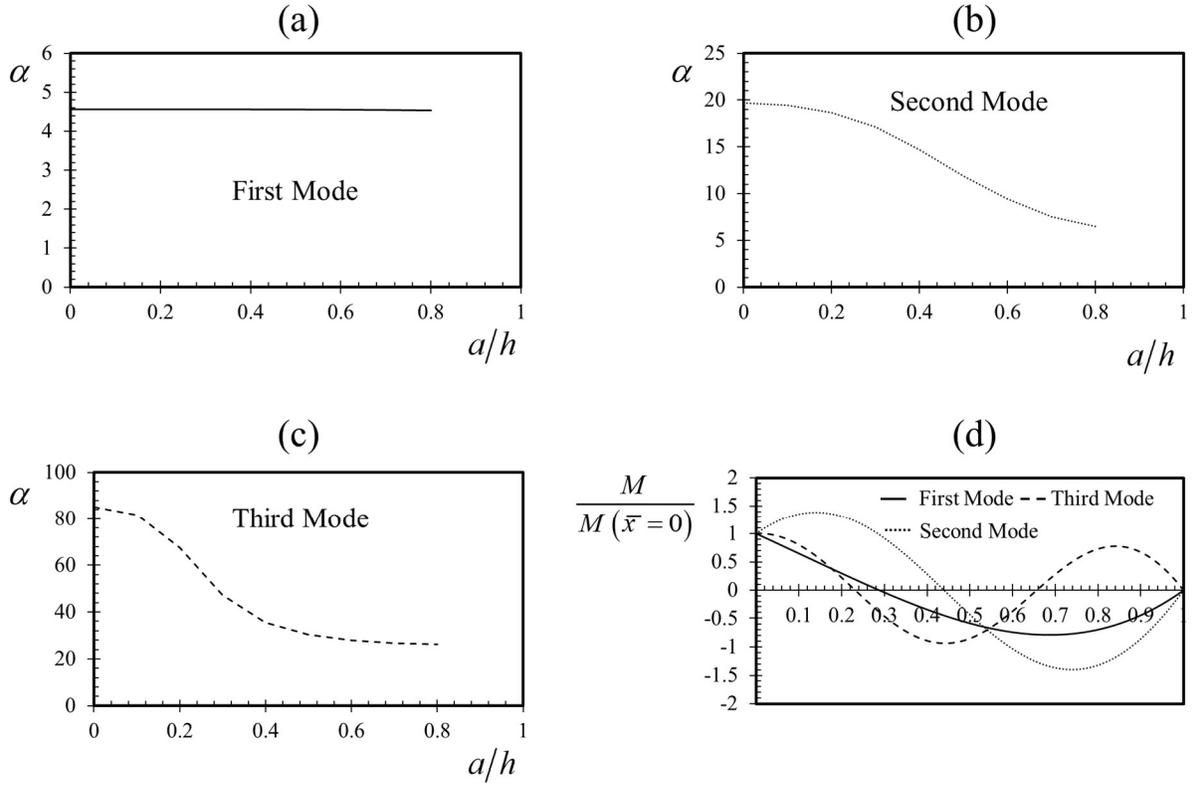

**Fig. 9** The critical loads corresponding to the (a) first, (b) second, and (c) third buckling modes of a clamped-free pillar with an elastic foundation with $\bar{K}_W = 300$ and $\bar{K}_P = 8$ by varying the crack length. The pillar has one crack at $\bar{x}_1 = 0.3$, and $h/L = 0.05$. (d) The axial variation of the normalized bending moment of a clamped-free pillar with no cracks.

*3.2.2 Effect of Crack Location*

It can be understood from Eq. (1) that the bending moment and shear force at the damaged cross-section dictate the impact of the damage on the buckling loads. Therefore, it is expected that changing the location of a crack significantly affects the buckling loads. This premise is validated by the results depicted in Fig. 10, which illustrate the buckling loads of pillars without elastic foundation containing cracks of length $a/h = 0.5$ as function of the crack location. The buckling loads are normalized with respect to the solutions for the intact pillars. The results are presented for different nonlocal parameters and boundary conditions.

In the considered case, the effect of the crack is mostly controlled by the bending moment at the damaged cross-section. More precisely, the cracks located at the cross-sections with higher bending moments have a stronger effect on the buckling load. This is the reason why the effect of the crack closer to the fixed end in the clamped-free pillar is higher than that of the cracks at further distances. For the pinned-pinned and clamped-clamped pillars, the buckling load is highly affected by the crack because the bending moment attains its maximum value at the location of the damaged cross-section that is the mid-length. The maximum effect of the crack on the buckling load of the clamped-pinned pillar occurs at a location within the interval $0.6 < \bar{x}_1 < 0.8$ depending on the nonlocal parameter.

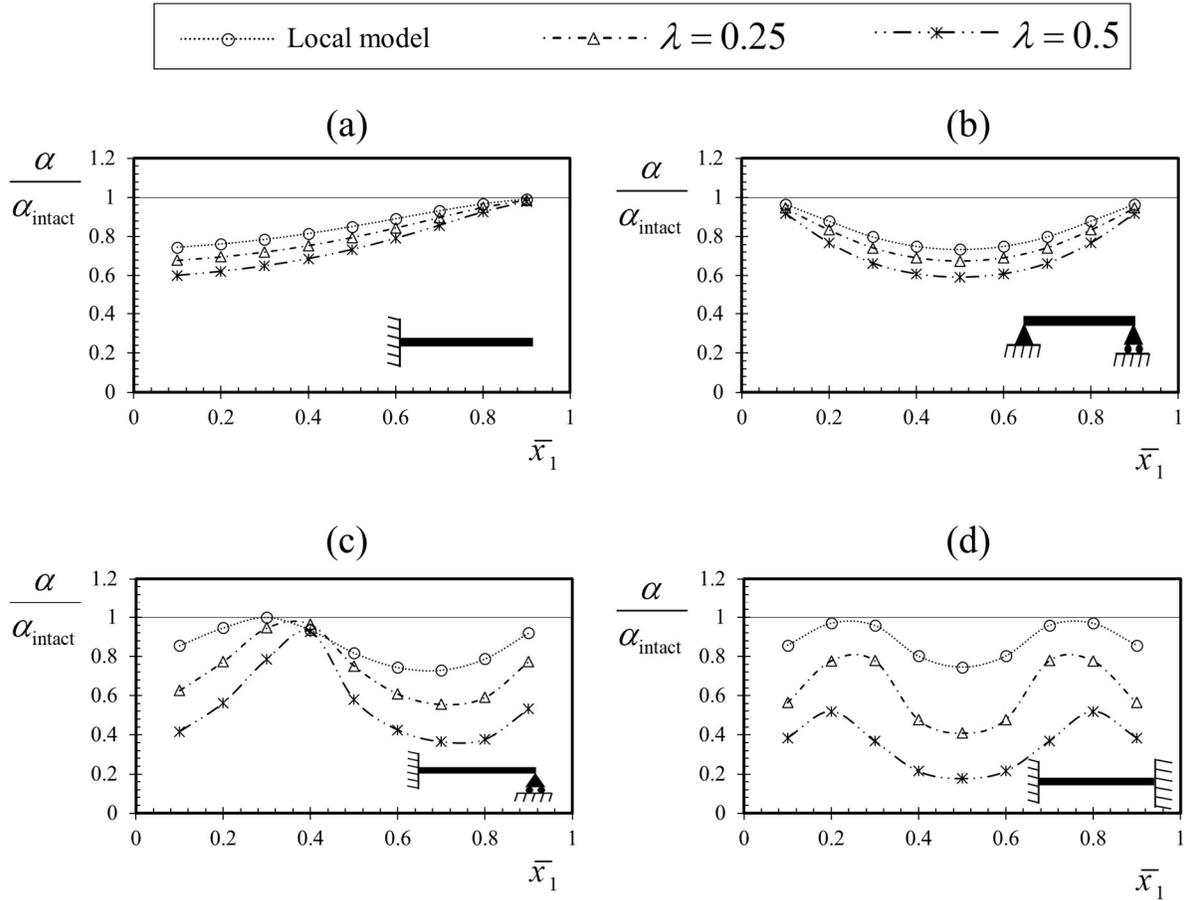

**Fig. 10** The buckling loads vs. the crack location and nonlocality for (a) clamped-free, (b) pinned-pinned, (c) clamped-pinned, and (d) clamped-clamped pillars in the absence of elastic foundation. Pillars have one crack with $a/h = 0.5$, and $h/L = 0.05$.

Since the axial variation of the bending moment depends on the buckling mode, it is expected that the impact of the damage location on the critical loads associated with the higher modes of buckling is different from that observed in Fig. 10. This is shown in Fig. 11 for the second mode of buckling of pillars without elastic foundation. The results in Fig. 11(a) and (b) refer to the second buckling loads of the clamped-free and pinned-pinned pillars by altering the location of the damaged cross-section and the nonlocality. The pillar containing a single crack of length $a/h = 0.3$ has the thickness to length ratio of $h/L = 0.05$. It can be understood from the comparison of Fig. 11 with Fig. 10 that the dependence of the second buckling load on the crack location is different from that of the first buckling mode.

The axial variation of the bending moments of the intact clamped-free and pinned-pinned pillars for $\lambda = 0.5$ are shown in Fig. 11(c) and (d). For the clamped-free pillar, the bending moment of the second buckling mode decreases from $\bar{x} = 0$ to approximately 0.4. This is the main reason why the effect of the crack on the second buckling load of the clamped-free pillar decreases for $0 < \bar{x} < 0.4$ (see Fig. 11(a)). The influence of the crack on the buckling load is found to be more significant when it occurs at cross-sections where the bending moment is higher. For example, the bending moment of the second buckling mode of the pinned-pinned pillar increases from $\bar{x} = 0$ to approximately 0.2 (Fig. 11(d)), resulting in a stronger impact of the damage on the critical loads (Fig. 11(b)).

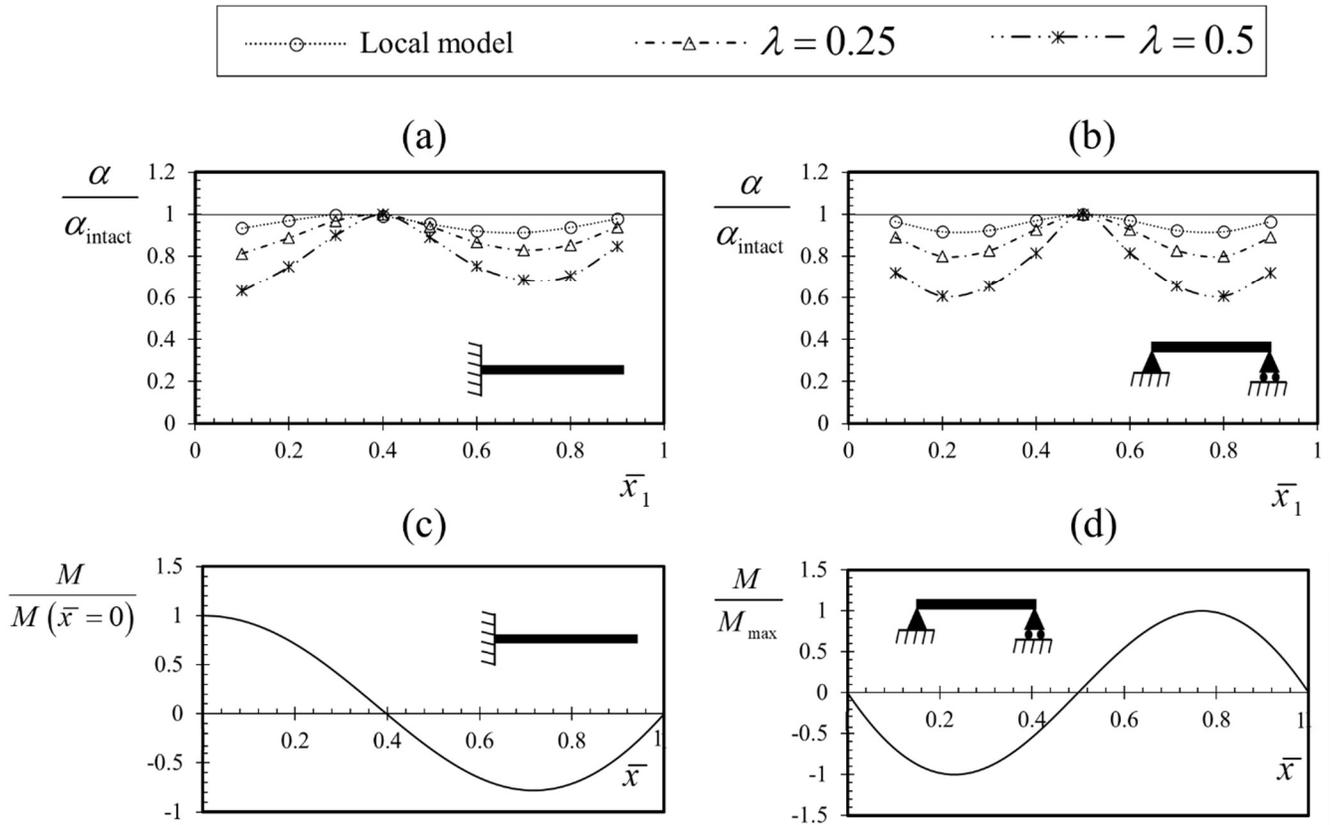

**Fig. 11** The second buckling loads of (a) clamped-free, and (b) pinned-pinned pillars without elastic foundation by altering the damage location and the nonlocality. Pillars have one crack with $a/h = 0.3$ and $h/L = 0.05$. The axial variation of bending moment for the (c) clamped-free and (d) pinned-pinned pillars free of cracks for $\lambda = 0.5$.

*3.2.3 Effect of Translational Spring*

To show that the influence of the translational spring related to the shear force at the cracked cross-section can be important in some cases, the first and higher critical loads are illustrated in Fig. 12 for a clamped-clamped pillar without elastic foundation. The pillar has one crack at $\bar{x}_1 = 0.6$, $h/L = 0.1$ and $\lambda = 0.5$. The buckling loads are demonstrated for two scenarios: (i) including the translational spring at the damaged cross-section, and (ii) excluding it. The latter refers to the crack compatibility conditions given in Eq. (1) for $c_V = 0$, which ensures that the deflection is continuous at the damaged cross-section. The first and second buckling loads do not considerably change when the contribution of the translational spring is excluded. However, for the third mode of buckling of pillars with long cracks, the exclusion of the translational spring from the model may result in a considerable overestimation of the buckling loads.

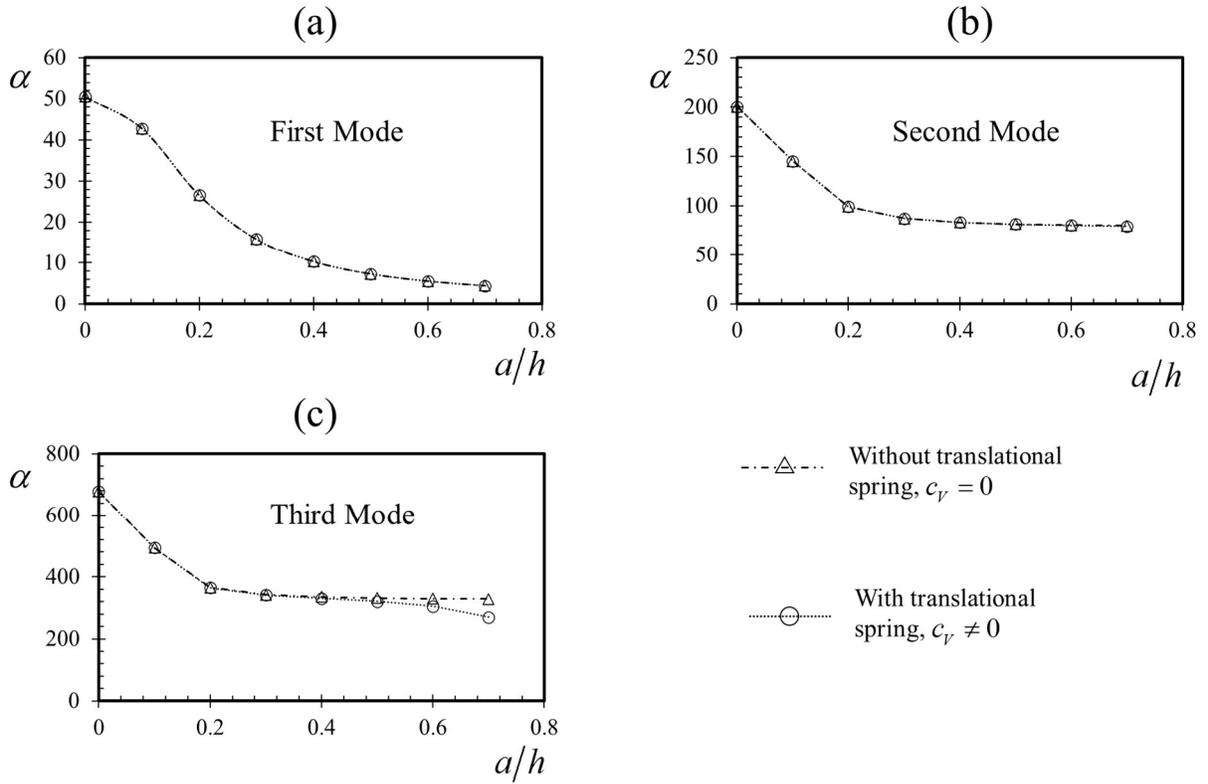

**Fig. 12** The critical loads of the (a) first, (b) second, and (c) third buckling modes of a clamped-clamped pillars without the elastic foundation in the case when the translational spring at the damaged cross-section is present (circles) or absent (triangles). The pillar contains a single crack at $\bar{x}_1 = 0.6$, and has $h/L = 0.1$ and $\lambda = 0.5$.

### 3.3. Miniaturized Pillars with Multiple Cracks

The present model is used to study the critical loads and associated buckling configurations of the pillars with two, three, and four cracks. As shown in Fig. 13, the first example is a clamped-free pillar with two cracks at $\bar{x}_1 = 0.1$ and $\bar{x}_2 = 0.3$. The buckling loads are calculated for the pillar with $h/L = 0.05$ in the absence of the elastic foundation. The length of the first crack is fixed at a constant value of $a_1/h = 0.5$, and the critical loads are illustrated in the figure by increasing the length of the second crack for different values of the nonlocal parameter. Results are normalized against the buckling loads of the pillar with only the second crack.

For the cases with smaller values of $a_2/h$, the presence of multiple cracks has an amplifying effect. This is because the buckling load of the pillar with two cracks is considerably lower than the buckling load of the pillar with either of the cracks. As $a_2/h$ increases, the amplification effect decreases. When the second crack is long enough, the effect of the first crack on the buckling load becomes negligible, and the buckling load of the structure tends to approach that of the pillar with only the second crack (i.e., shielding effect). The nonlocality intensifies and reduces, respectively, the amplification and shielding effects.

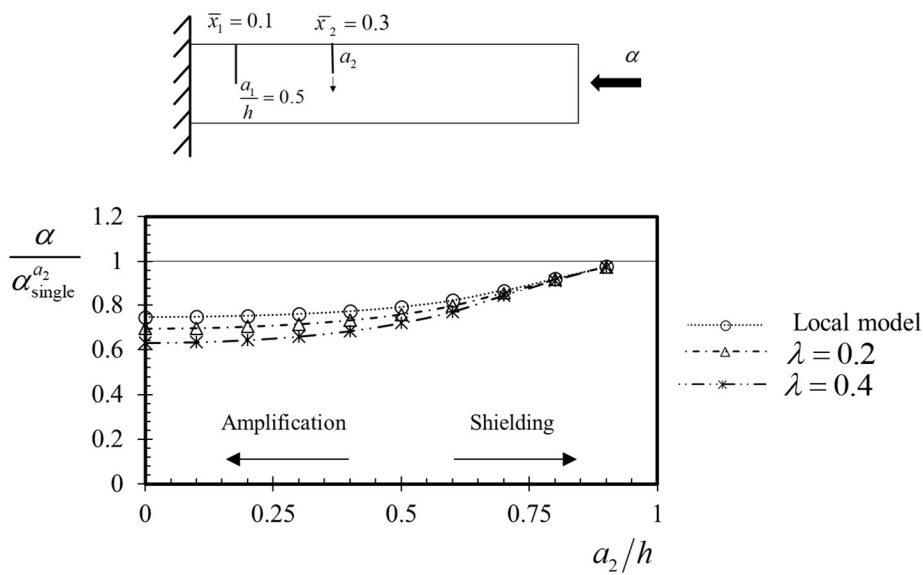

**Fig. 13** The buckling loads of a clamped-free pillar without elastic foundation with two damaged cross-sections at $\bar{x}_1 = 0.1$ and $\bar{x}_2 = 0.3$, $a_1/h = 0.5$, and $h/L = 0.05$ by varying $a_2/h$ and the nonlocal parameter.

Cracks may also significantly change the buckling mode shapes of the miniaturized pillars. The mode shapes corresponding to the first mode of buckling of the clamped-free, pinned-pinned, clamped-pinned, and clamped-clamped miniaturized intact (solid lines) and cracked (dotted lines) pillars without elastic foundation are presented in Fig. 14 for $h/L = 0.05$ and different values of the nonlocal parameter. The cracked pillars have two cracks at $\bar{x}_1 = 0.2$ and $\bar{x}_2 = 0.6$. The crack lengths of the clamped-free and pinned-pinned pillars are $a_1/h = a_2/h = 0.5$, and those of the clamped-pinned and clamped-clamped pillars are $a_1/h = a_2/h = 0.25$. For the local pillar, the cracks slightly affect the mode shapes of the pinned-pinned and clamped-clamped pillars, and have almost no effect on those of the clamped-free and clamped-pinned pillars. However, the mode shapes of the pinned-pinned, clamped-pinned, and clamped-clamped nonlocal pillars are strongly affected by the presence of cracks. The change in the mode shape is more pronounced for the clamped-clamped boundary condition. The mode shape of the nonlocal clamped-free pillar remains almost unaffected by the presence of the two cracks.

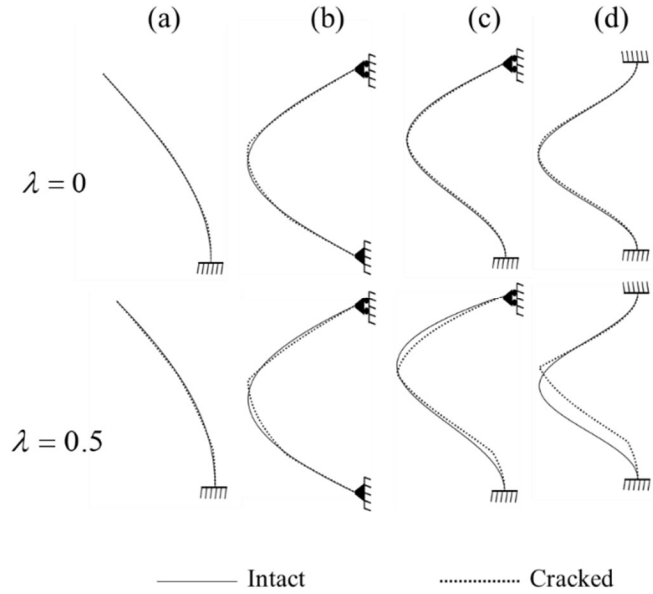

**Fig. 14** The mode shapes of (a) clamped-free, (b) pinned-pinned, (c) clamped-pinned, and (d) clamped-clamped miniaturized intact (solid lines) and cracked (dotted lines) pillars without elastic foundation. The mode shapes are illustrated for $h/L = 0.05$ and $\lambda = 0$ and $0.5$. The damaged pillars have cracks at $\bar{x}_1 = 0.2$ and $\bar{x}_2 = 0.6$. The crack lengths of the clamped-free and pinned-pinned pillars are $a_1/h = a_2/h = 0.5$, and those of the clamped-pinned and clamped-clamped pillars are $a_1/h = a_2/h = 0.25$.

The critical loads of clamped-clamped pillars without elastic foundation with three and four cracks are presented in Fig. 15. The cracks have the same lengths, $a_i/h = \bar{a}$ for $i = 1, \ldots, 3$, or $4$, and are located across the pillar with equal spacing. The critical loads are depicted for $h/L = 0.05$ and $\lambda = 0.5$ by varying the crack length. It can be seen in the figure that the pillars with longer cracks have lower buckling loads. The buckling load of the pillar with four cracks is always lower than that of the pillar with three cracks, regardless of the crack length. This is because the stiffness of the pillar with four cracks is lower than that of the pillar with three cracks.

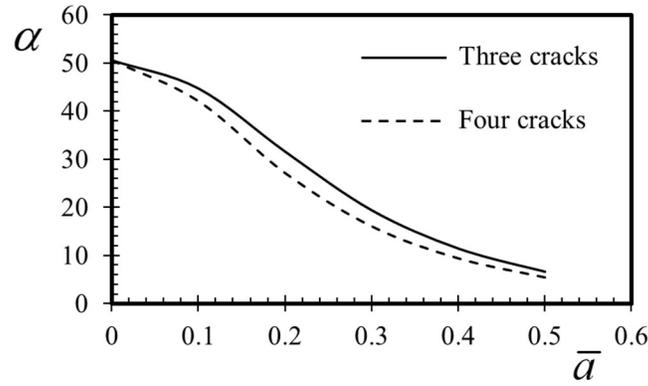

**Fig. 15** The critical loads of clamped-clamped pillars with three and four cracks having the same length $\bar{a}$ and equal spacing by altering the crack severity in the absence of the elastic foundation. The critical loads are depicted for $h/L = 0.05$ and $\lambda = 0.5$.

## 4. Conclusions

The effects of multiple edge cracks, shear force at the damaged cross-section, elastic foundation, and type of boundary conditions on the critical loads and buckling configurations of the small-scale pillars have been studied. The formulation combines the Bernoulli-Euler beam model with the stress-driven nonlocal theory. The sections at the left and right of a damaged cross-section have been assumed to be interconnected by rotational and translational springs.

Pillars with typical boundary conditions and with one to four edge cracks have been considered. The buckling loads have been presented by varying the crack length, crack location, nonlocality, and stiffness of the elastic foundation. The present work has yielded the following conclusions:

- The presence of edge cracks has a more pronounced impact on the critical loads and buckling configurations of miniaturized pillars compared to their larger counterparts.
- The shielding and amplification effects due to the presence of multiple cracks are more significant in the buckling behavior of miniaturized pillars than large-scale pillars.
- The effect of translational spring related to the shear force at the damaged cross-section should not be excluded from the formulation when dealing with higher modes of buckling and long cracks.
- Elastic foundation decreases the effect of the cracks on the buckling loads.
- Depending on the crack's location and the boundary conditions, the critical loads can be greatly influenced by the presence of a crack.

**Funding**

The first author gratefully acknowledges the financial support provided by the National Science Centre (NCN) in Poland through the grant agreement No: UMO-2022/47/D/ST8/01348. For the purpose of Open Access, the authors have applied a CC-BY public copyright license to any Author Accepted Manuscript (AAM) version arising from this submission.

**Statements and Declarations**

The authors declare that they have no known competing financial interests or personal relationships that could have appeared to influence the work reported in this paper.

## Appendix A. Crack Compliances

The compliances of the rotational and translational springs in Eq. (1) are [33]:

$$_i c_M = 12Ch \int_0^{\zeta_i} F_M^2(\zeta_i) d\zeta_i \quad \text{for } 0 \leq \zeta_i \leq 0.6$$

$$F_M(\zeta_i) = \sqrt{\tan \frac{\pi \zeta_i}{2}} \frac{0.923 + 0.199\left[1 - \sin \frac{\pi \zeta_i}{2}\right]^4}{\cos \frac{\pi \zeta_i}{2}} \quad (14)$$

$$_i c_M = 2.65335 Ch \int_0^{\zeta_i} \frac{1}{(1-\zeta_i)^3} d\zeta_i \quad \text{for } 0.6 < \zeta_i < 1$$

and

$$_i c_V = \frac{1}{6} Ch^3 \int_0^{\zeta_i} \frac{1}{1-\zeta_i} F_V^2(\zeta_i) d\zeta_i \quad \text{for } 0 \leq \zeta_i < 1$$

$$F_V(\zeta_i) = 1.993 \zeta_i + 4.513 \zeta_i^2 - 9.516 \zeta_i^3 + 4.482 \zeta_i^4 \quad (15)$$

where $\zeta_i = a_i/h$. The expressions in Eqs. (14) and (15) can be derived through the utilization of both energy considerations and linear elastic fracture mechanics principles.

## Appendix B. Boundary Conditions

The boundary conditions for four typical cases are:

**Table B1:** Variational boundary conditions.

| Boundary type | Condition | Dimensionless condition |
|---|---|---|
| Clamped-Clamped | $^{(1)}v(0) = 0$ <br> $^{(1)}v_{,x}(0) = 0$ <br> $^{(n+1)}v(L) = 0$ <br> $^{(n+1)}v_{,x}(L) = 0$ | $^{(1)}\eta(0) = 0$ <br> $^{(1)}\eta_{,\bar{x}}(0) = 0$ <br> $^{(n+1)}\eta(1) = 0$ <br> $^{(n+1)}\eta_{,\bar{x}}(1) = 0$ |
| Clamped-Pinned | $^{(1)}v(0) = 0$ <br> $^{(1)}v_{,x}(0) = 0$ <br> $^{(n+1)}v(L) = 0$ <br> $^{(n+1)}v_{,xx}(L) - L_C^2\, ^{(n+1)}v_{,xxxx}(L) = 0$ | $^{(1)}\eta(0) = 0$ <br> $^{(1)}\eta_{,\bar{x}}(0) = 0$ <br> $^{(n+1)}\eta(1) = 0$ <br> $^{(n+1)}\eta_{,\bar{x}\bar{x}}(1) - \lambda^2\, ^{(n+1)}\eta_{,\bar{x}\bar{x}\bar{x}\bar{x}}(1) = 0$ |
| Pinned-Pinned | $^{(1)}v(0) = 0$ <br> $^{(1)}v_{,xx}(0) - L_C^2\, ^{(1)}v_{,xxxx}(0) = 0$ <br> $^{(n+1)}v(L) = 0$ <br> $^{(n+1)}v_{,xx}(L) - L_C^2\, ^{(n+1)}v_{,xxxx}(L) = 0$ | $^{(1)}\eta(0) = 0$ <br> $^{(1)}\eta_{,\bar{x}\bar{x}}(0) - \lambda^2\, ^{(1)}\eta_{,\bar{x}\bar{x}\bar{x}\bar{x}}(0) = 0$ <br> $^{(n+1)}\eta(1) = 0$ <br> $^{(n+1)}\eta_{,\bar{x}\bar{x}}(1) - \lambda^2\, ^{(n+1)}\eta_{,\bar{x}\bar{x}\bar{x}\bar{x}}(1) = 0$ |
| Clamped-Free | $^{(1)}v(0) = 0$ <br> $^{(1)}v_{,x}(0) = 0$ <br> $^{(n+1)}v_{,xx}(L) - L_C^2\, ^{(n+1)}v_{,xxxx}(L) = 0$ <br> $\frac{1}{C}\left[^{(n+1)}v_{,xxx}(L) - L_C^2\, ^{(n+1)}v_{,xxxxx}(L)\right] +$ <br> $+ (P - K_P)\, ^{(n+1)}v_{,x}(L) = 0$ | $^{(1)}\eta(0) = 0$ <br> $^{(1)}\eta_{,\bar{x}}(0) = 0$ <br> $^{(n+1)}\eta_{,\bar{x}\bar{x}}(1) - \lambda^2\, ^{(n+1)}\eta_{,\bar{x}\bar{x}\bar{x}\bar{x}}(1) = 0$ <br> $\left[^{(n+1)}\eta_{,\bar{x}\bar{x}\bar{x}}(1) - \lambda^2\, ^{(n+1)}\eta_{,\bar{x}\bar{x}\bar{x}\bar{x}\bar{x}}(1)\right] +$ <br> $+ \left(\alpha\pi^2 - \bar{K}_P\right)^{(n+1)}\eta_{,\bar{x}}(1) = 0$ |